\documentclass[%
 aip,
 amsmath,amssymb,
 graphicx,
 reprint]{revtex4-1}

\draft 

\usepackage{graphicx}
\usepackage{dcolumn}
\usepackage{bm}

\usepackage[utf8]{inputenc}
\usepackage[T1]{fontenc}
\usepackage{mathptmx}
\usepackage{etoolbox}

\makeatletter
\def\@email#1#2{%
 \endgroup
 \patchcmd{\titleblock@produce}
  {\frontmatter@RRAPformat}
  {\frontmatter@RRAPformat{\produce@RRAP{*#1\href{mailto:#2}{#2}}}\frontmatter@RRAPformat}
  {}{}
}%

\begin{document}


\title{A quantum cascade laser-pumped molecular laser tunable over 1 THz} 



\author{Arman Amirzhan}
\thanks{These authors contributed equally to this paper}
\author{Paul Chevalier}
\thanks{These authors contributed equally to this paper}
\affiliation{Harvard John A. Paulson School of Engineering and Applied Sciences, Harvard University, Cambridge, MA 02138, USA}
\author{Jeremy Rowlette}
\author{H. Ted Stinson}
\author{Michael Pushkarsky}
\author{Timothy Day}
\affiliation{DRS Daylight Solutions, San Diego, CA 92128, USA}
\author{Henry O. Everitt}
\affiliation{DEVCOM Army Research Lab, Houston, TX  77005, USA}
\affiliation{Department of Physics, Duke University, Durham, NC 27708, USA}
\author{Federico Capasso}
\email[Authors to whom correspondence should be addressed: ]{capasso@seas.harvard.edu (F.C) or  everitt@phy.duke.edu (H.O.E.)}
\affiliation{Harvard John A. Paulson School of Engineering and Applied Sciences, Harvard University, Cambridge, MA 02138, USA}




\date{\today}

\begin{abstract}
Despite decades of research, no frequency tunable sources span the terahertz gap between 0.3 - 3 THz.  By introducing methyl fluoride (CH$_3$F) as a new gain medium for a quantum cascade laser-pumped molecular laser (QPML), we demonstrate continuous-wave lasing from more than 120 discrete transitions spanning the  range from 0.25  to 1.3 THz. Thanks to its large permanent dipole moment and large rotational constants, methyl fluoride (CH$_3$F) as a QPML gain medium combines a lower threshold, larger power efficiency, and wider tuning range than other molecules. These key features of the CH$_3$F QPML, operated in a compact cavity at room temperature, pave the way to a versatile THz source to bridge the THz gap.
\end{abstract}


\maketitle 

As terahertz (THz) technologies mature and create exciting applications in radio astronomy~\cite{wootten2009atacama}, biomedicine~\cite{sun2017recent}, communications, and the defense/aerospace industries ~\cite{ellrich2020terahertz}, the generation and detection of THz radiation remains a challenge. Despite decades of extensive research in the area, few practical sources of THz radiation exist. Vacuum electronic devices and multiplier chain sources struggle to reach frequencies above 1 THz, free electron lasers and optically pumped far infrared (IR) lasers (OPFIR) are bulky and non-portable, and difference frequency photomixers produce broad linewidths and low output power~\cite{brundermann2012sources}. THz quantum cascade lasers, despite recent advances, can operate only in pulsed mode using thermoelectric coolers~\cite{khalatpour2021high} and require cryogenic cooling for cw operation~\cite{wienold2014high}.

A new source of widely tunable THz radiation has been recently demonstrated: the Quantum Cascade Laser (QCL) pumped molecular laser (QPML)~\cite{pagies2016low,chevalier2019widely,wienold2020laser}. The QPML, a new class of OPFIR lasers~\cite{chang1970laser,everitt1986dynamics}, requires the pumping of a mid-IR ro-vibrational transition with an external laser to create population inversions in dipole-allowed transitions between adjacent rotational energy levels, with lasing frequencies in the THz range. The advantage of the QPML concept is that it uses a compact, continuously-tunable mid-IR QCL as a pumping source instead of a bulky, line-tunable mid-IR gas laser. The latter requires a frequency coincidence between a CO$_2$ laser line and a mid-IR ro-vibrational transition to produce lasing on one or two, often unfavorable, THz rotational transitions. To change OPFIR laser frequency, the user has to change the molecular gain medium and the pump laser line.  By contrast, the continuous tunability of QCLs allows $any$ ro-vibrational transition to be pumped for $any$ gas phase molecules with a permanent electric dipole moment, so the challenge is to identify the ideal molecular gain medium that provides the best combination of power and tunability. Then, placing this molecular gain medium in a small laser cavity~\cite{everitt1986dynamics, wang2018high} will create a room temperature cw laser in a compact form factor with the potential for wide frequency tunability throughout the THz region. 

We have previously demonstrated the widely tunable QPML concept by using an external cavity QCL as a pump source and nitrous oxide (N$_2$O) as a molecular gain medium, achieving low power continuous-wave laser emission for 39 transitions with discrete line tunability from 0.25 to 0.95 THz~\cite{chevalier2019widely}.  The first demonstration of a QPML used ammonia gas molecules (NH$_3$) as a gain medium~\cite{pagies2016low} and was based on population inversion of pure inversion transitions in the $\nu_2=1$ excited vibration band. Such pure inversion transitions can achieve higher output power, with more than 1 mW demonstrated recently~\cite{lampin2020quantum}, but with a tuning range limited to a few frequencies between 0.9 to 1.1 THz. Another QPML demonstration using ammonia produced five more lasing lines in the range from 4.4 to 4.5 THz using direct rotational transitions in the same $\nu_2=1$ excited vibration ~\cite{wienold2020laser}.

Here we introduce methyl fluoride (CH$_3$F) as the most promising gas-phase QPML molecular gain medium to date, by demonstrating lasing from more than 120 of the 315 possible lines, spanning the frequency range 0.25 - 1.3 THz, always producing at least 10 times more output power than from the N$_2$O QPML operating at the nearest emission frequency. Laser emission beyond 1 THz is reliably achieved thanks to larger rotational constants and lower lasing threshold at the optimal working pressure. The low lasing threshold and the high power efficiency of the CH$_3$F QPML permitted by the molecule's large permanent dipole moment makes it suitable for applications even with the QCL pump emitting as little as 50 mW. Thanks to its compactness, efficiency, wide tuning range, and room temperature operation, the methyl fluoride QPML has the potential to become a key technology to bridge the THz gap from 200 GHz up to 2 THz for applications in imaging, security, or communications.

\begin{figure*}[htbp]
\centering
\includegraphics[width=0.75\linewidth]{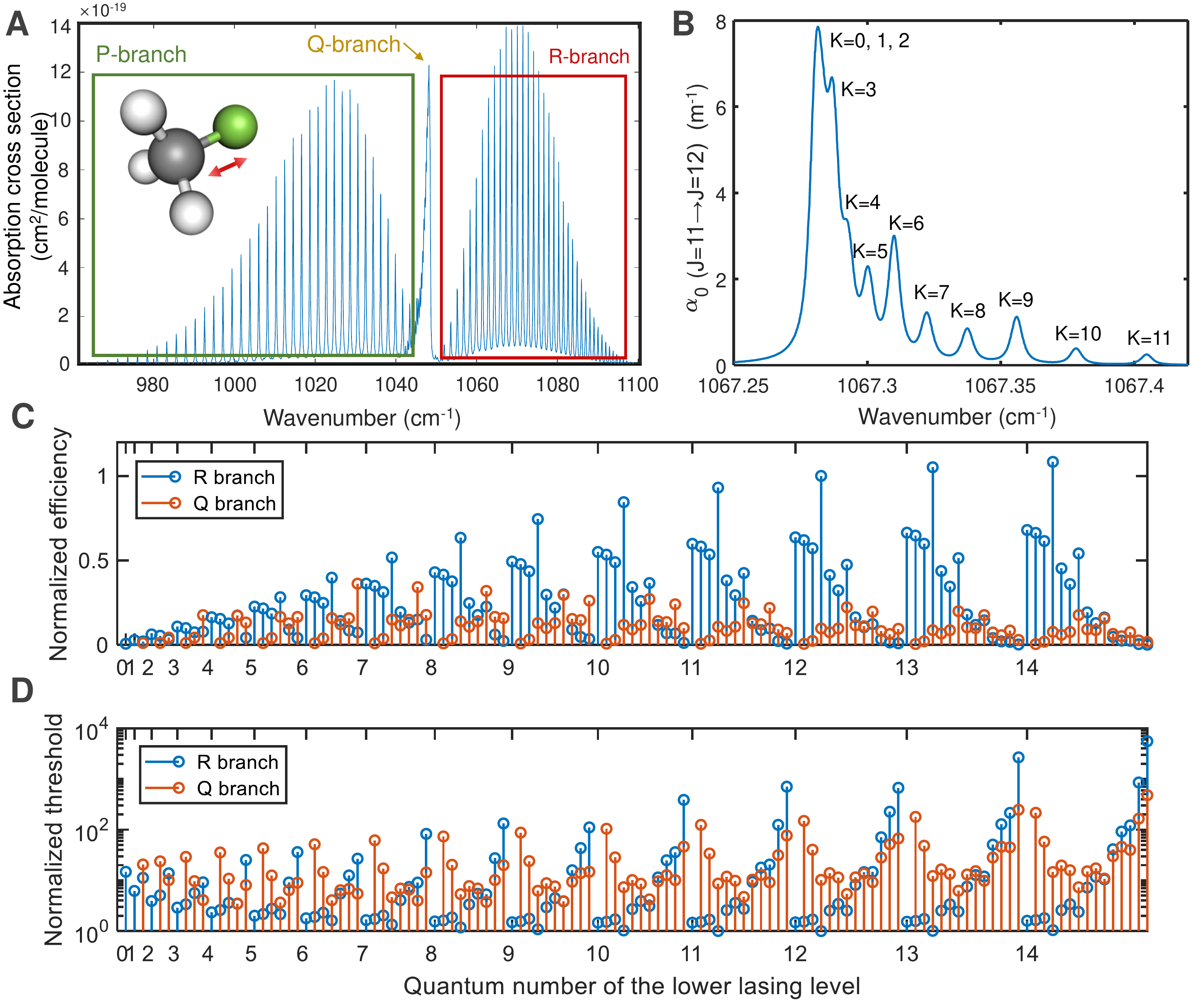}
\caption{(A) Room temperature IR spectrum of the $\nu_3=1$ ro-vibrational band of CH$_3$F around 1000 cm$^{-1}$, showing the absorption cross sections as a function of wavenumber for the P, Q, and R branch transitions, using data obtained from the HITRAN database\cite{HITRAN2016_XSC,HITRAN-1164}. (B) Simulated room temperature spectrum of the Doppler broadened R-branch ro-vibrational transitions from the $J_L=11$ ground vibrational level at 30~mTorr, using Ref.~\citenum{phillips2012infrared}.
Estimated relative power efficiency (C) and threshold (D) plotted as a function of lower laser level $J_U-1$ and $K$, normalized by the respective values for the $J=13 \to 12$, $K=3$ transition. For a Q-branch pump $J_U-1=J_L-1$, and for an R-branch pump $J_U-1=J_L$. }
\label{fig:fig1}
\end{figure*}

Methyl fluoride is a prolate symmetric top molecule (see Fig.~\ref{fig:fig1}(A) inset) whose ro-vibrational energy level structure is characterized by two quantum numbers: $J$, representing the total angular momentum, and $K$, representing the projection of the angular momentum along the main symmetry axis. The energies of the rotational states within a given vibrational energy band, up to the 4-th order in $J$ and $K$, are given by~\cite{townes1975microwave}

\begin{equation}
\begin{split}
    F(J,K)= & B J(J+1)+(A-B) K^2 \\
    & -D_J J^2(J+1)^2 - D_{JK} J (J+1) K^2 - D_K K^4,
    \label{eq:eq1}
\end{split}
\end{equation}
where $B$ and $A$ are rotational constants, and $D_J$, $D_{JK}$, and $D_K$ are centrifugal distortion constants, unique for each vibrational energy band. A more thorough polynomial development is given in the supplementary information (SI).  As methyl fluoride is a prolate symmetric top molecule, its rotational constant $A$ ($\approx  150$ GHz) is larger than $B$ ( $\approx 25$ GHz) \cite{phillips2012infrared}, so the energy of rotational states increases with increasing $J$ and $K$. Given the dipole-allowed $\Delta J=0, \pm 1, \Delta K = 0$ transition selection rules~\cite{townes1975microwave}, the laser transition frequency between adjacent rotational energy states can be calculated using equation~\ref{eq:eq1} to be:

\begin{equation}
\begin{split}
    \nu(J,K) & = F(J+1,K)-F(J,K) \\ 
    & = 2 B(J+1) - 4D_J (J+1)^3 - 2D_{JK} (J+1) K^2
    \label{eq:eq2}
    \end{split}
\end{equation}

Our QCL frequency can be tuned across the lowest  energy $\nu_3$ vibrational band of CH$_3$F \cite{phillips2012infrared}, whose IR spectrum is shown in Fig.~\ref{fig:fig1}(A) and whose carbon-fluorine stretching is illustrated in inset. This molecular gain medium is pumped from a ground state rotational level with quantum numbers $J_L$, $K_L$ to a rotational state in the $\nu_3=1$ vibrational level with quantum numbers $J_U$, $K_U$. The CH$_3$F gain medium was pumped either with an R-branch transition ($J_U=J_L+1$, $K=K_U=K_L$) or a Q- branch transition ($J_U=J_L$, $K=K_U=K_L\neq0$). P-branch IR transitions could have been pumped too, but they are so similar to R-branch transitions, except for their frequency, that they were not considered here. The number of allowed IR transitions between a given $J_L \to J_U$ increases with increasing $J_L$, spanning $K = 0, 1,... J_L$ with frequencies that separate quadratically with $K$ because of centrifugal distortion (equation~\ref{eq:eq2}), as can be seen in Fig.~\ref{fig:fig1}(B). Because the IR branching ratio transition matrix element is proportional to $\frac{(J+1)^2 -K^2}{(J+1)(2J+1)}$ for R-branch transitions and to $\frac{K^2}{J(J+1)}$ for Q-branch transitions~\cite{townes1975microwave}; R-branch pumping favors low $K$ transitions and Q-branch pumping is more effective for high $K$ transitions.

A rotational population inversion can occur within the $\nu_3=1$ vibrational state between levels with quantum numbers $J_U$ and $J_U-1$. Because the emission frequency from equation~\ref{eq:eq2} depends on $K$, for a given $J_U$ there are $J_U$ possible laser transitions ($K = 0$ to $J_U-1$) for R-branch pumping and $J_U-1$ ($K = 1$ to $J_U-1$) for Q-branch pumping. It is this increasing number of discrete laser lines with increasing $J_U$, coupled with the ability to pump either R- or Q-branch transitions with a single QCL to maximize power, that makes CH$_3$F an attractive gain medium for high tunability, achieved by selectively pumping ro-vibrational transitions of different $K$ values.

The THz emission power and lasing threshold of the CH$_3$F QPML, which can be estimated using a comprehensive model~\cite{wang2018high} that considers all molecular relaxation mechanisms, can be reasonably approximated by a far simpler model~\cite{chevalier2019widely, wang2021optimizing} in the low pressure regime where the rate of unfavorable molecular dipole-dipole collisions is smaller than the rate of favorable hard collisions with the cavity walls. The salient expressions, recalled in the SI, indicate that at a given frequency the output power increases and the lasing threshold decreases with increasing absorption strength of the pumped IR transition. 

Although actual laser power and threshold sensitively depend on cavity geometry and loss, the simple model thereby indicates that one may ascertain the relative efficacy of the allowed laser transitions when pumped by P, Q, or R-branch transitions by comparing their relative IR absorption strengths. For thermally populated rotational energy levels ($i.e. F(J,K) < kT$) in $\nu_0$, these can be easily calculated (see SI) using equation~\ref{eq:eq1} and the known degeneracies to estimate the fractional occupation of $J_L$, $K_L$, multiplied by the branching ratio for the considered IR pumping branch (P, Q, or R). Likewise, the relative power efficiency may be estimated by multiplying the relative IR absorption by $J_U$ to account for the frequency dependence of the laser transition, while the relative lasing threshold may be estimated by multiplying the relative IR absorption by the branching ratio for the lasing transition. 

For the CH$_3$F QPML, the efficiency and the lasing threshold for many of these lines are compared for both  R- and Q-branch pumping in Figs.~\ref{fig:fig1}(C) and~\ref{fig:fig1}(D), respectively, as a function of the lower lasing level quantum numbers $J_U-1$. For a Q-branch pump $J_U-1=J_L-1$, and for an R-branch pump $J_U-1=J_L$. The plots in Figs.~\ref{fig:fig1}(C) and~\ref{fig:fig1}(D) have been respectively normalized to the efficiency and threshold of the lasing line with lowest lasing threshold ($J = 13 \to 12$, $K=3$).  Notice that the lowest thresholds generally occur for low $K$ lines when pumped by R (and P) branch transitions but for mid-range $K$ lines when pumped by Q branch transitions. Likewise, the highest power efficiencies generally occur for $K=3$ lines when pumped by R (or P) branch transitions and $K=6$ lines for Q branch pumping. This simple analysis provides an easy way to compare the performance of the many allowed laser transitions from a QPML with a symmetric-top molecular gain medium, allowing one to ascertain the most efficient and the lowest threshold lasing lines.

\begin{figure}[htbp]
\centering
\includegraphics[width=\linewidth]{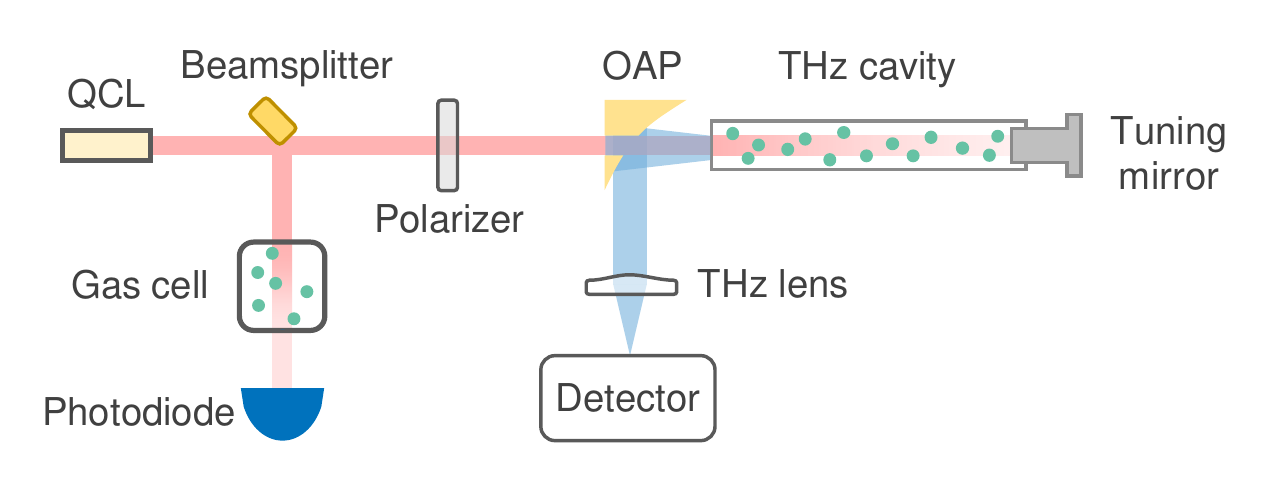}
\caption{Schematic of the experimental setup. A gold-plated silicon wafer was used as a beamsplitter to reflect a small portion (about 5\% of the incoming power) of the pump beam into the reference gas cell, while the rest entered the THz cavity. The following detectors were used to detect lasing lines: Schottky diodes between 0.25 and 1.1 THz, a Golay cell above 1.1 THz, and heterodyne receivers to measure spectra between 0.3 and 1.1 THz.}
\label{fig:fig2}
\end{figure}

Figure~\ref{fig:fig2} shows a simplified schematic of the experimental setup, a more detailed schematic is shown in SI. The IR pump is provided by an external cavity (EC)-QCL (Daylight Solutions 41095-HHG-UT, tunable from 920 to 1194 cm$^{-1}$) emitting up to 250 mW. The THz cavity is a copper pipe with a 4.8 mm internal diameter and 50 cm in length. A flat mirror with a centered 1 mm diameter pinhole was used as the output coupler. The cavity resonance frequency was tuned by changing the tuning mirror position to adjust the cavity length. A separate CH$_3$F gas absorption cell was used to monitor how precisely the QCL emission frequency was tuned to the desired CH$_3$F ro-vibrational transition. About 5\% of the incoming power was sampled and redirected towards this gas cell by using a small piece of a gold-coated silicon wafer that partially clipped the edge of the QCL beam.

The QCL pump beam was coupled into the THz laser cavity through the pinhole in the output coupler using an anti-reflective coated ZnSe lens to focus the beam into the pinhole. A 2 mm thick ZnSe window mounted at the Brewster angle was used to maximize the amount of QCL power injected into the vacuum-tight gas laser cavity. A wire grid polarizer mounted on a rotational stage was used as a variable attenuator for the linearly polarized QCL beam. The injected IR power was measured by replacing the QPML rear tuning mirror with a ZnSe window and measuring the pump beam power transmitted through the THz cavity. After passing through the coupling optics, the maximum power transmitted through the cavity was measured to be about 150 mW at 1065 cm$^{-1}$.

\begin{figure*}[htbp]
\centering
\includegraphics[width=0.75\linewidth]{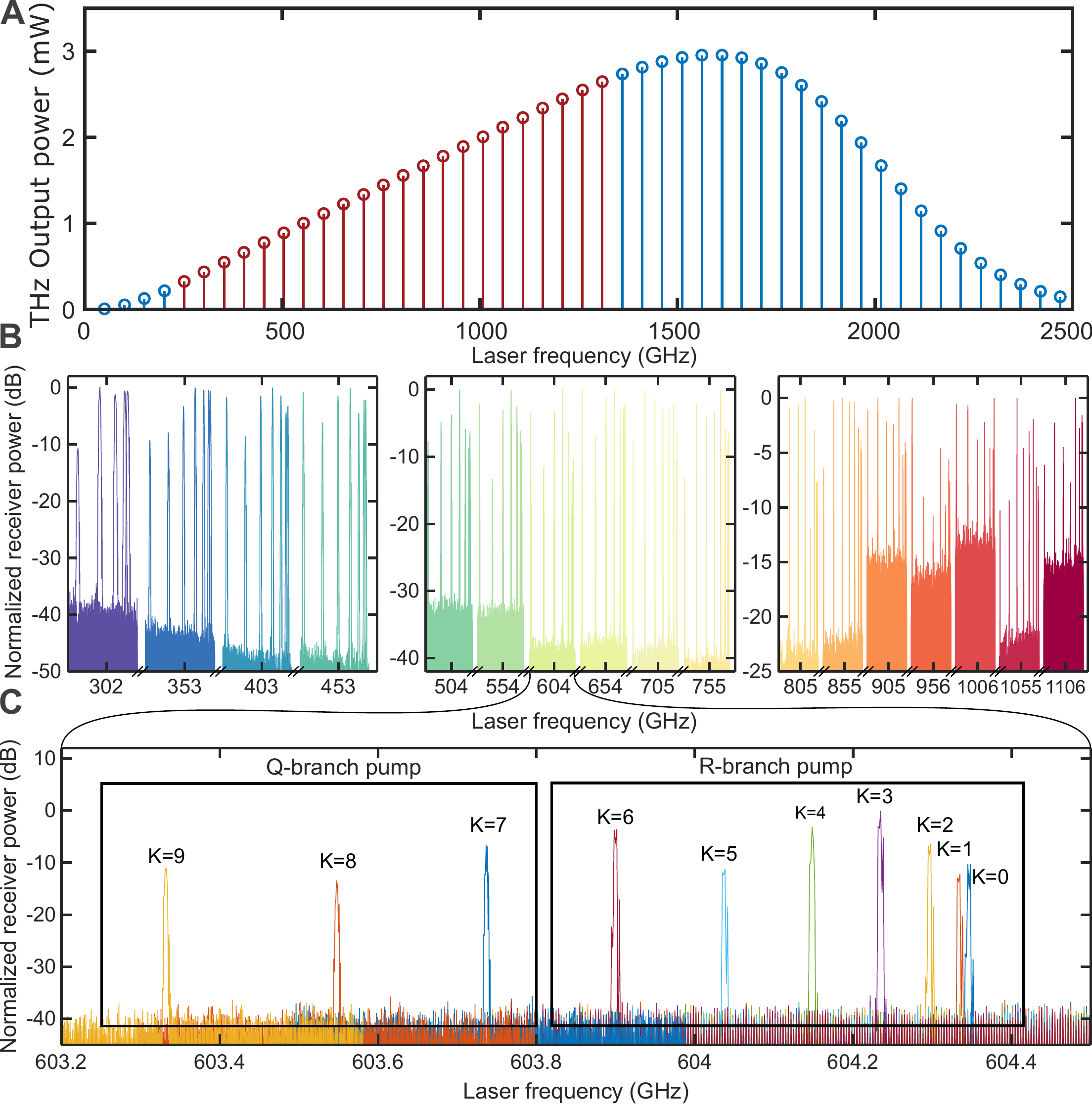}
\caption{(A) Plot of estimated maximum THz emission power using a simple model~\cite{wang2021optimizing} for a 30 mTorr CH$_3$F QPML pumped with a 150 mW QCL, as a function of the frequency for a 4.8 mm diameter, 50 cm long cell. Red lines represent transitions at which laser emission was attempted and achieved. Blue lines represent transitions at which laser emission is theoretically possible but was either not achieved or not attempted. (B) Experimental data showing all lines measured using a heterodyne receiver, from 302 GHz to 1106 GHz. (C) Detailed view of measured lasing lines emitting around 603 GHz showing which lines were pumped with an R- or Q-branch transition.}
\label{fig:fig3}
\end{figure*}

The off-axis parabolic mirror (OAP) used to collect the generated THz laser radiation had a hole drilled along the focal axis through which the pump beam was coupled into the THz laser cavity. A polytetrafluoroethylene (PTFE) lens blocked any residual IR signal and focused the collected THz beam onto a Schottky diode detector, a Golay cell, or a heterodyne receiver. The THz output power was measured using a calibrated calorimeter-style power meter (VDI PM5B).

The theoretically predicted output power of a CH$_3$F QPML pumped by a 150 mW QCL with a gas pressure of 30 mTorr is plotted in Fig.~\ref{fig:fig3}(A). This plot was obtained using the enhanced simple model previously mentioned~\cite{wang2021optimizing} that takes into account dipole-dipole collisions, multiple passes of the infrared pump, but for that particular plot no infrared saturation was considered. The output power was calculated as a function of the laser frequency, assuming that the gain medium was pumped using an R-branch transition with $J_L=1$ to 50, and $K=3$ (except for $J_L \leq$ 3 where $K=0$ was assumed). Maximum QPML power is expected between 1 to 1.7 THz.

Laser emission was obtained experimentally by tuning the QCL into coincidence with a Q- or R-branch ro-vibrational transition of the gain medium, and then by tuning the cavity resonance into coincidence with the lasing rotational transition by adjusting the cavity length with a differential micrometer screw (Newport DM17-4, 100 nm fine sensitivity) of a translation stage. Emission was achieved on lines between 0.25 THz to 1.3 THz by pumping ro-vibrational transitions of CH$_3$F originating from a level with quantum number $J_L$ spanning from 5 to 25, and various $K$ numbers. Of the 315 possible lasing lines over this range, pumping was attempted and emission was observed for at least 120 of them (shown as red lines on Fig.~\ref{fig:fig3}(A)), including many measured using a heterodyne receiver that are plotted together in Fig.~\ref{fig:fig3}(B). The blue lines in Fig.~\ref{fig:fig3}(A) represent transitions for which laser emission is possible but was either not attempted (for frequencies above 1.3 THz) or was not possible in our laser cavity (for frequencies below 250 GHz). In Fig.~\ref{fig:fig3}(C), a detailed view of lasing lines obtained around 604 GHz is shown, indicating which lines were pumped by an R-branch transition or a Q-branch transition. As anticipated, lines with larger $K$ quantum number were more efficiently pumped with Q-branch transitions, while the lines with lower $K$ quantum number were more efficiently pumped with R-branch transitions. Because the lower $K$ lines are spectrally close together, we were routinely able to induce two, three, even four transitions to lase simultaneously with R-branch pumping.

The output power of the THz laser emission depends on a combination of many parameters, including the choice of the gain medium, the IR pumping efficiency, and the laser cavity design~\cite{wang2021optimizing}. The choice of the gain medium will be motivated by many properties, including its IR absorption coefficient and saturation, the  dipole moments of the lasing transitions, the dipole-dipole collision rate, and the ratio of the emission frequency to the pump frequency. The pumping efficiency will be primarily affected by the available QCL power and the overlap between the QCL emission linewidth and the gain medium absorption bandwidth. The effects of the laser cavity are determined by its geometry, cavity quality factor, loss factor of pump power inside the cavity, and the THz losses in the output coupler. All the above factors must be considered in calculating the optimum THz cavity length, diameter, output coupling ratio, and gas pressure~\cite{wang2021optimizing}. Increasing the cavity diameter results in improved cavity Q factor and reduced pump saturation, but at the expense of increased inversion quenching caused by the reduced relaxation rate through wall collisions. Optimal cavity dimensions, operating pressure, and more accurate output powers are calculated by performing extensive molecular dynamics simulations~\cite{wang2018high,wang2021optimizing}.

\begin{figure}[htbp]
\centering
\includegraphics[width=\linewidth]{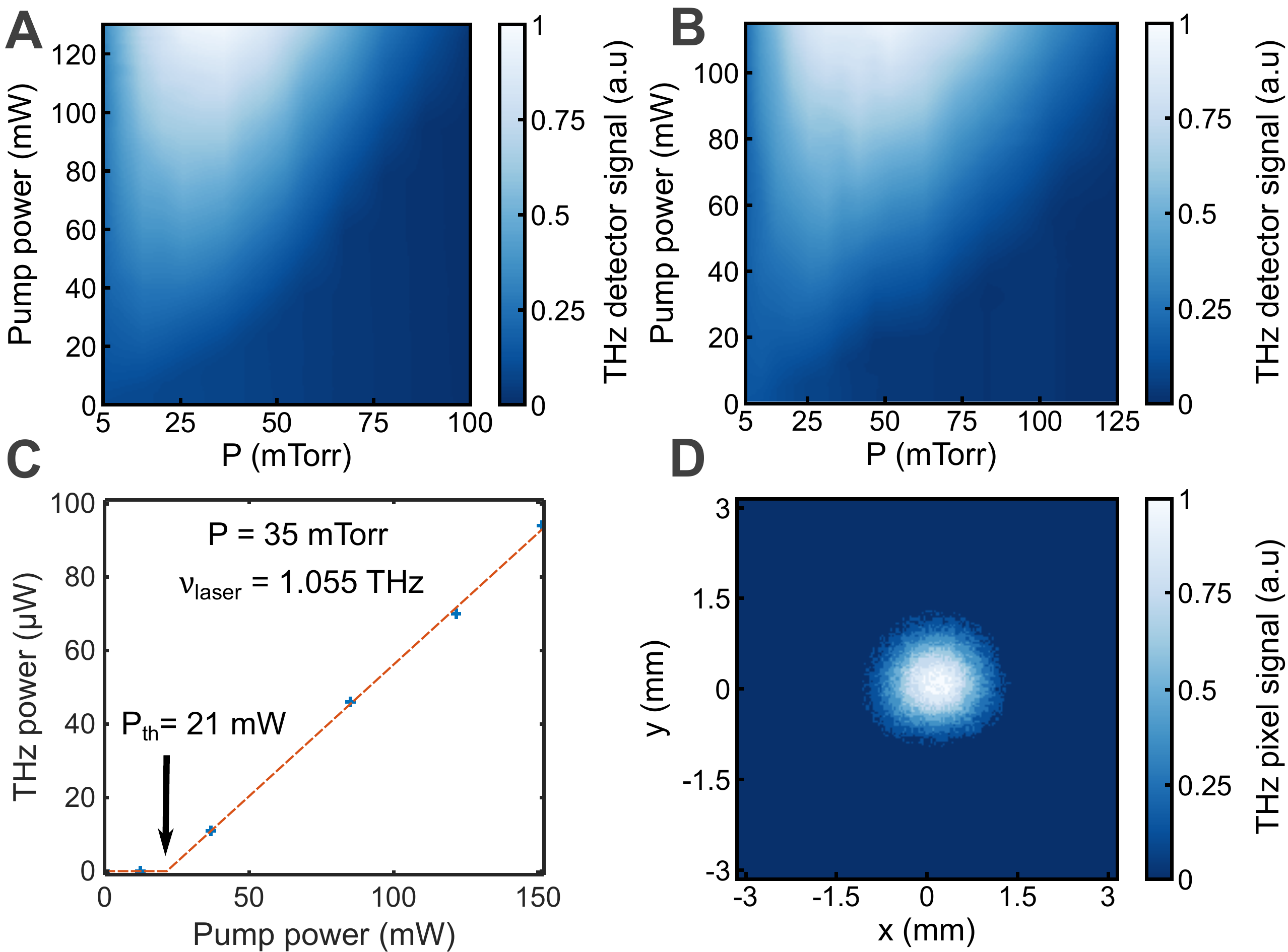}
\caption{Experimental data showing the detected THz power as a function of gas pressure and IR pump power for emission near (A) 604 GHz and (B) 1.1 THz. (C) Output power vs. pump power curve obtained with a gas pressure of 35 mTorr for the $J=21 \to 20$, $K=3$ transition at 1.055 THz. (D) Profile of the THz focused spot for the same transition measured with a microbolometer camera.}
\label{fig:fig4}
\end{figure}

The laser performance has been measured experimentally at various gas pressures and pump powers. The experiment was conducted by setting the pressure to a fixed value, then slowly adjusting the QCL pump power while measuring the THz signal strength using a Schottky diode detector. By referencing the measured signal strength against the largest measured power for a given lasing transition, a surface plot of normalized THz output signal as a function of pump power and pressure can be constructed, such as shown in Fig.~\ref{fig:fig4}(A) for the $J=12 \to 11$, $K=3$ lasing transition near 604 GHz and in Fig.~\ref{fig:fig4}(B) for the $J=22 \to 21$, $K=3$ lasing transition near 1.1 THz. The threshold increases with increasing pressure because of the pressure-dependent dipole-dipole collision rate that operates to quench the inversion~\cite{wang2021optimizing}. The pressure for optimal power increases with increasing pump power as expected, but it also increases with increasing THz emission frequency, from $\sim$~30 mTorr for 604~GHz to $\sim$~50 mTorr for 1.1~THz. This occurs because above J$_L$ = 11 the fractional population decreases with increasing $J_L$, thereby requiring more pressure to avoid pump saturation and achieving comparable IR absorption strength.

Comparing different laser transitions, the largest THz output power occurs when the highest population inversion is obtained, determined primarily by the absorption strength of the pumped transition and the lasing frequency to pump frequency ratio. Consequently, the output power will vary significantly with THz emission frequency, as shown in Figure~\ref{fig:fig3}(A). The largest measured THz power (13 \textmu W) was obtained for the $J=12 \to 11$, $K=3$ transition near 604 GHz, for a pump power of 145 mW at a pressure of 30 mTorr. Because the QPML is increasingly sensitive to cavity misalignment with increasing frequency, more power was measured at 604 GHz than at 1.055 THz, a problem that will be overcome with improved cavity design. About 9.5 \textmu W of collected power was measured at around 1.055 THz (for the $J=21 \to 20$, $K=3$ transition) for a pump power of 150 mW at a pressure of 35 mTorr. 

The laser power collection efficiency through the ZnSe Brewster window, the off-axis parabolic mirror, and the PTFE lens was experimentally estimated at most at 10\%, which corresponds to a total emitted power of at least 130 \textmu W at 604 GHz and at least 95 \textmu W at 1.055 THz. The normalized power efficiency (defined by the power efficiency divided by the ratio of the emitted photon energy by the pump photon energy) is 4.7\% at 604 GHz and 2\% at 1.055 THz. These power efficiencies compare favorably with the performance of commercial OPFIR lasers, and greater efficiency approaching the theoretical maximum of 50\% is possible with improved cavity design and alignment~\cite{wang2021optimizing}. 

The estimated emitted THz power as a function of pump power for the 1.055 THz transition is plotted in Fig.~\ref{fig:fig4}(C), where the collected power was measured for five different pump power values (blue crosses); the red dashed line is a linear fit of the THz power above threshold. A clear threshold of around $21~$mW pump power can be seen. The profile of the laser spot, as focused by the PTFE lens for the same transition, was measured using a microbolometer camera (INO Microxcam 384i-THz) and is shown in Fig.~\ref{fig:fig4}(D). The spot was fitted with a two-dimensional Gaussian model with spot sizes $2w_x= 2.08~$mm and $2w_y= 1.89~$ mm.

In summary, we have shown that by using a widely tunable external cavity quantum cascade laser to pump gaseous methyl fluoride, a prolate symmetric top molecule with a large dipole moment and rotational constant, created a widely tunable THz lasing over many lasing lines can be achieved. It is this wide tunability, coupled with its low threshold, ability to pump either R- or Q-branch transitions to maximize power, good conversion efficiency, high operating frequency, and compact size, that make QPML's in general, and the CH$_3$F QPML in particular, a leap-ahead source technology for generating THz radiation~\cite{pagies2016low,chevalier2019widely,wienold2020laser}.

\section*{Supplementary Material}
See Supplement 1 for additional details about the experimental and theoretical work presented in this manuscript.

\begin{acknowledgments}
The external cavity QCL used in the experiments was provided by DRS Daylight Solutions. The authors acknowledge Stanley Cotreau and Andrew DiMambro of Harvard University Instructional machine shop for their help with fabrication of the THz cavity elements. This work was partially supported by the U.S. Army Research Office (contracts W911NF-19-2-0168, W911NF-20-1-0157). Any opinions, findings, conclusions, or recommendations expressed in this material are those of the authors and do not necessarily reflect the views of the Assistant Secretary of Defense for Research.
\end{acknowledgments}

\section*{Data Availability Statement}
The data that support the findings of this study are available from the corresponding author upon reasonable request.

\section*{References}
\bibliography{references}

\end{document}



\title{A quantum cascade laser-pumped molecular laser tunable over 1 THz: supporting information} 



\author{Arman Amirzhan}
\thanks{These authors contributed equally to this paper}
\author{Paul Chevalier}
\thanks{These authors contributed equally to this paper}
\affiliation{Harvard John A. Paulson School of Engineering and Applied Sciences, Harvard University, Cambridge, MA 02138, USA}
\author{Jeremy Rowlette}
\author{H. Ted Stinson}
\author{Michael Pushkarsky}
\author{Timothy Day}
\affiliation{DRS Daylight Solutions, San Diego, CA 92128, USA}
\author{Henry O. Everitt}
\affiliation{DEVCOM Army Research Lab, Houston, TX  77005, USA}
\affiliation{Department of Physics, Duke University, Durham, NC 27708, USA}

\author{Federico Capasso}
\email[Corresponding authors: ]{capasso@seas.harvard.edu (F.C)  or  everitt@phy.duke.edu (H.O.E.)}
\affiliation{Harvard John A. Paulson School of Engineering and Applied Sciences, Harvard University, Cambridge, MA 02138, USA}




\date{\today}

\begin{abstract}
Despite decades of research, no frequency tunable sources span the terahertz gap between 0.3 - 3 THz.  By introducing methyl fluoride (CH$_3$F) as a new gain medium for a quantum cascade laser-pumped molecular laser (QPML), we demonstrate continuous-wave lasing from more than 120 discrete transitions spanning the  range from 0.25  to 1.3 THz. Thanks to its large permanent dipole moment and large rotational constants, methyl fluoride (CH$_3$F) as a QPML gain medium combines a lower threshold, larger power efficiency, and wider tuning range than other molecules. These key features of the CH$_3$F QPML, operated in a compact cavity at room temperature, pave the way to a versatile THz source to bridge the THz gap.
\end{abstract}

\pacs{}

\maketitle 

\section{Energy levels of a symmetric top molecule}

\begin{figure}[htbp]
\centering
\includegraphics[width=\linewidth]{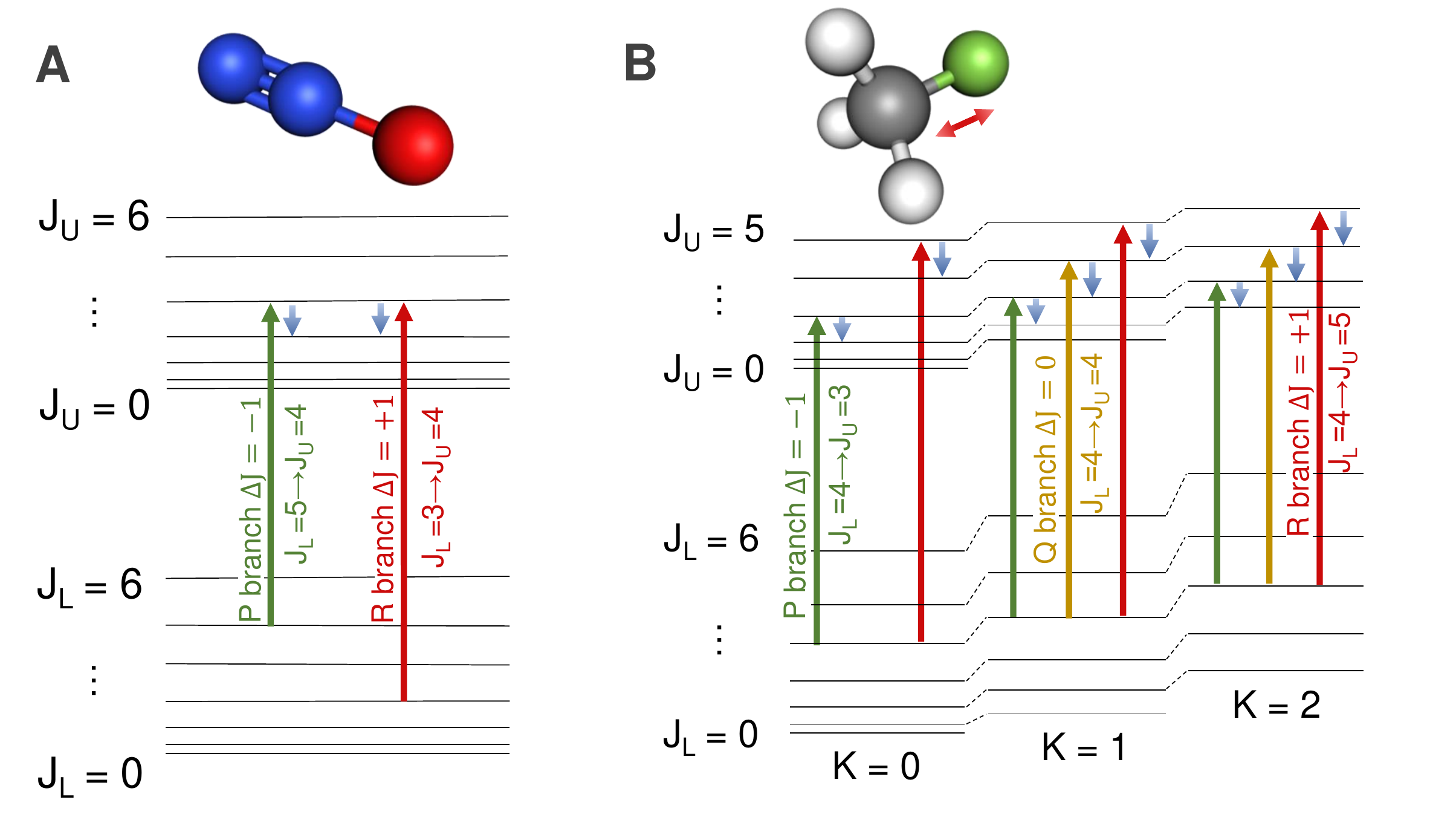}
\caption{ (A) Diagram of the rotational energy levels in the ground (labeled $J_L$) and excited (labeled $J_U$) vibrational states of  a linear molecule (e.g. N$_2$O). A three-dimensional schematic of the N$_2$O molecule is given in inset. (B) Diagram of the rotational energy levels in the ground state ($J_L$,$K$) and excited vibrational state ($J_U$,$K$) for a symmetric top molecule (e.g. CH$_3$F). A three-dimensional schematic of the CH$_3$F molecule is given in inset with a red arrow indicating the $v_3$ carbon-fluorine stretching vibrational mode considered in this paper. The green, yellow and red arrows represent respectively P, Q, and R branch infrared transitions, while the blue arrows represent the corresponding lasing transitions.}
\label{fig:fig1}
\end{figure}

In our previously reported QPML laser, we used nitrous oxide (N$_2$O), a linear molecule with only one unique non-zero moment of inertia. Such a molecule has a simple roto-vibrational spectrum with only one quantum number ($J$) defining rotational energy levels (see Fig.~\ref{fig:fig1}(A)). Methyl fluoride (also known as fluoromethane or CH$_3$F), on the other hand, is a prolate symmetric top molecule (see Fig.~\ref{fig:fig1}B inset). Compared to the linear rotor, the symmetric top geometry has a more complex roto-vibrational energy level structure, with two quantum numbers defining the molecule's rotational energy levels, as shown in Fig.~\ref{fig:fig1}(B). The first quantum number, $J$, represents the total angular momentum, while the second quantum number $K$ defines the projection of the  angular momentum along the main symmetry axis.
The selection rules prohibit a change in $K$ quantum number, while the change in $J$ number $\Delta J$ must be 0, -1 or 1\cite{townes1975microwave}. In Fig.~\ref{fig:fig1}(B), red arrows identify R-branch infrared transitions initiating from the same $J_L$ quantum number but with different $K$ quantum numbers (respectively green and yellow arrows for P and Q branch transitions). Q-branch transitions cannot occur between $K=0$ levels for the $v_3$ vibrational mode (and all other symmetrical modes) of CH$_3$F. Each infrared transition corresponding to different $K$ values will have a unique frequency, which can be seen as absorption peaks in Fig.~1(B) of the main text. Pumping each transition will create population inversion between rotational energy levels, leading to stimulated emission with different emission frequencies, which are shown as blue arrows in Fig.~\ref{fig:fig1}(B). Therefore, it is possible to have finite discrete tunability of the THz emission frequency for a transition between two given $J$ quantum numbers by selectively pumping roto-vibrational transitions of different $K$ values.

The Equation 1 and 2 of the main text simply approximate the rotational energy levels of the CH$_3$F molecule. For the sake of simplicity we only introduced in the main text the development up to the 4-th order in powers of $J$ and $K$.
However the rotational constants for methyl fluoride have been experimentally fitted up to the 8-th order in power of $J$ and $K$~\cite{phillips2012infrared}. Sextic and hexic constants are required to properly calculate the vibrational transition energy for the molecules. The following equation is used to calculate energies of rotational states within a vibrational energy band up to the 8-th order in $J$ and $K$\cite{townes1975microwave,phillips2012infrared}:

\begin{equation}
\begin{split}
    F(J,K)= & B J(J+1)+(A-B) K^2 \\
    & -D_J J^2(J+1)^2 - D_{JK} J (J+1) K^2 - D_K K^4 \\
    &+ H_J J^3(J+1)^3 + H_{JJK} J^2(J+1)^2 K^2 + H_{JKK} J(J+1) K^4 + H_K K^6 \\
    &+ L_J J^4(J+1)^4 + L_{JJJK} J^3(J+1)^3 K^2 + L_{JJKK} J^2(J+1)^2 K^4 + L_{JKKK} J(J+1) K^6  + L_K K^8
    \label{eq:eq1}
\end{split}
\end{equation}
where $J$ and $K$ are rotational quantum numbers, $B$ and $A$ are rotational constants, $D_J$, $D_{JK}$, and $D_K$, $H_J$, $H_{JJK}$, $H_{JKK}$, $H_K$, $L_J$, $L_{JJJK}$, $L_{JJKK}$,$L_{JKKK}$ and $L_K$ are centrifugal distortion constants. Rotational and centrifugal distortion constants are unique for each vibrational energy band. As methyl fluoride is a prolate symmetric top molecule its rotational constant $A$ ($\approx  150$ GHz) is always larger than $B$ ( $\approx 25$ GHz) \cite{phillips2012infrared}. The rotational constants for Methyl-Fluoride and used for theoretical calculations in this work are reproduced in Table~\ref{tab:tabConstants}. This results in increasing energy of rotational states with increasing $K$. The energy difference between adjacent rotational energy states defines the output THz lasing frequency and can be calculated using equation~\ref{eq:eq2} up to the 8-th order in $J$ and $K$. The dominant term for small values of $K$ corresponds to a quadratic decrease of the lasing frequency with increasing $K$ as can be seen further in Fig.~2(C) of the manuscript.

\begin{equation}
\begin{split}
    & F(J+1,K)- F(J,K)= 2 B(J+1) - 4D_J (J+1)^3 - 2D_{JK}  (J+1) K^2 \\
    &+ H_J (J+1)^3 ((J+2)^3-J^3) + 4H_{JJK}  (J+1)^3 K^2 + 2H_{JKK} (J+1) K^4  \\
    &+ L_J (J+1)^4 ((J+2)^4-J^4) + 6L_{JJJK}  (J+1)^3 ((J+2)^3-J^3) K^2 \\ 
    &+  4L_{JJKK}  (J+1)^3 K^4 + 2L_{JKKK}  (J+1) K^6
    \label{eq:eq2}
    \end{split}
\end{equation}

\begin{table*}[htp]
\caption{\label{tab:tabConstants} Rotational constants for methyl fluoride ($^{12}$CH$_3$F), recopied from Ref.\citenum{phillips2012infrared}}
\begin{tabular*}{0.4\textwidth}{c|c|c}
\hline\hline
& $\nu_3=0$ & $\nu_3=1$ \\
\hline
$A$ (MHz) & 155 352.70 & 155 058.474 \\
$B$ (MHz) & 25 536.14980 & 25 197.51060 \\
$D_j$ (kHz) & 60.2214 & 56.8868 \\ 
$D_{jk}$ (kHz) & 439.8156 & 518.2386 \\
$D_{k}$ (kHz) & 2108.0&2014.9000\\
$H_{j}$ (Hz) & -0.03214&-0.19320\\
$H_{jjk}$ (Hz) & 1.9470& 16.07690\\
$H_{jkk}$ (Hz) & 22.3610 & -96.44947\\
$H_{k}$ (Hz) & 0 &0.1197871 \\
$L_{j}$ (mHz) & 0 & 0.01020\\
$L_{jjjk}$ (mHz) & 0 & -1.7468\\
$L_{jjkk}$ (mHz) & 0 & 29.5648\\
$L_{jkkk}$ (mHz) & 0 & -173.2365\\
$L_{k}$ (mHz) & 0 & 0\\
$E_0$ (cm$^{-1}$)& - & 1048.610701\\
\hline
\end{tabular*}
\end{table*}

\section{Details on the laser operation}

\begin{figure}[htbp]
\centering
\includegraphics[width=\linewidth]{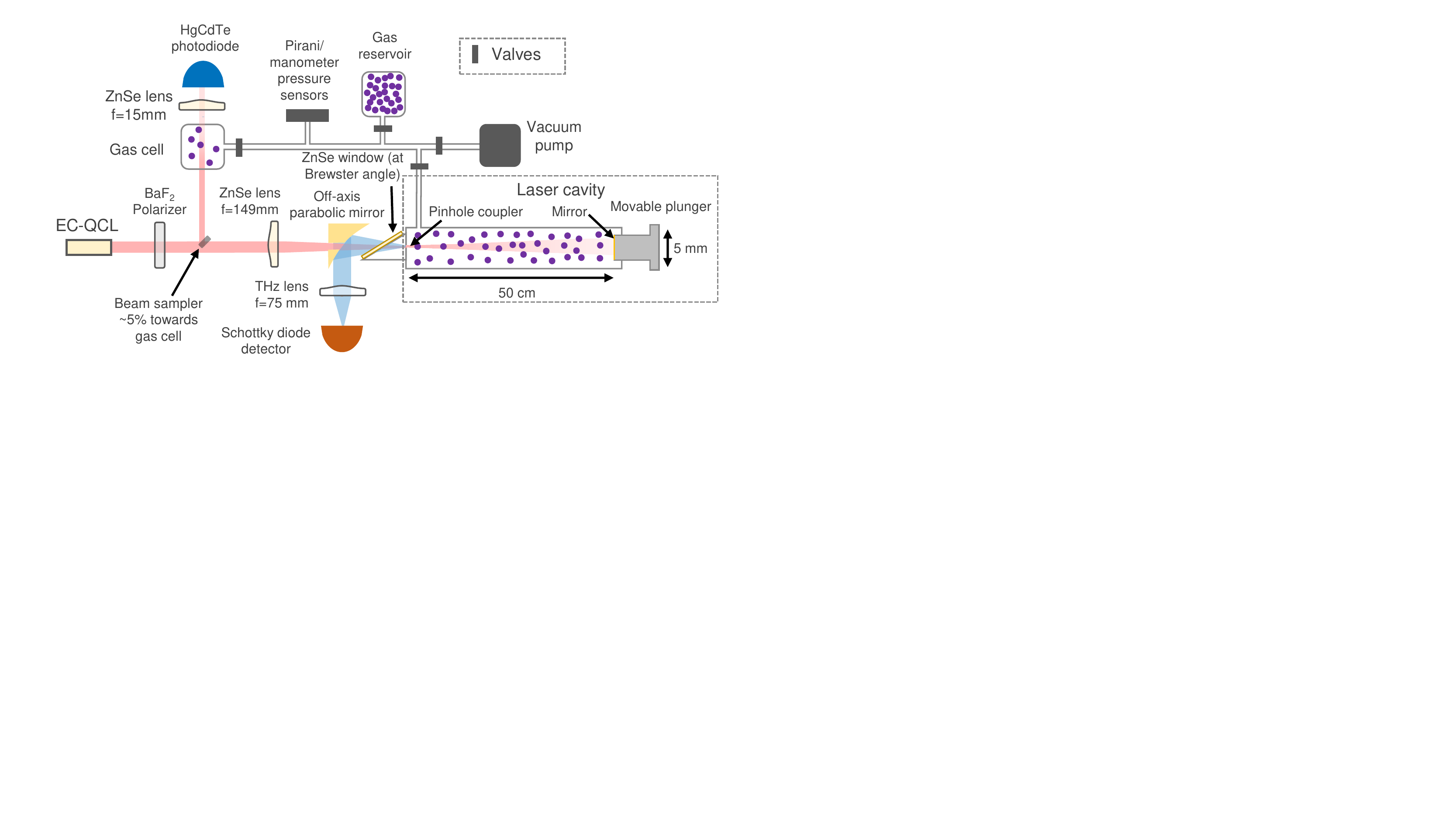}
\caption{Complete schematic of the experimental setup: light from the QCL passes through a polarization (used as a variable attenuator), a 149 mm focal length ZnSe lens, a hole through an off-axis parabolic mirror and then the Brewster window of the laser cavity.}
\label{fig:figSetup}
\end{figure}

A more complete schematic of the experimental setup than the one shown in Fig. 2 of the manuscript is provided in Fig.~\ref{fig:figSetup}. This experimental setup identifies the Brewster window used on the front of the laser cavity along with the vacuum pumps, valves and pressure sensors. The laser cell components (cavity, windows etc.) are sealed using Viton orings.

The pressure in the laser cell is controlled similar to the way described in our previous work~\cite{chevalier2019widely}, recycling the gas by cryogenic pumping in order to  minimize waste. First, the cell is evacuated to a very high vacuum ($P<10^{-5}$~mTorr) by means of a rough rotary vane pump (Varian DS102) and a turbo molecular pump (Varian Turbo V81). Gas phase methyl fluoride is contained in a small 10 mL metallic cylinder connected to the system with a needle valve. This cylinder is cooled down by liquid nitrogen which causes the gas to solidify. After this initial cool down, the valve to the vacuum pump is closed and the valve of the cylinder is opened. Then the liquid nitrogen is removed and the cylinder is allowed to warm, which causes the gas to sublime and fill the rest of the system. The valve of the cylinder is closed when the proper pressure is achieved. The pressure in the cell is measured by a thermocouple Gauge (Alcatel sensor AP1004 and controller ACR1000) and an absolute manometer (MKS Bar-a-tron, range 1 Torr). To remove gas from the system, the cylinder is cooled again and the needle valve is opened until the gas is recovered and the laser system is empty. 

In order to achieve laser emission, first the laser cavity is filled with CH$_3F$ at the desired pressure, then the QCL is tuned to a desired gas absorption line by setting the correct drive current, grating position, and laser temperature in the QCL controller. The precise drive conditions of the QCL are achieved by monitoring the transmission of the QCL through a reference gas cell containing 50 mTorr of CH$_3$F.

Finally, laser emission at THz frequencies is obtained by tuning the length of the cavity into resonance with the transition frequency and monitoring the emitted radiation with a Schottky diode detector (VDI ZBD WR2.2, WR1.5, WR1.2, WR1.0 or WR0.65 depending on the emission frequency) or a Golay Cell. The emission frequency was measured with a heterodyne mixer (VDI MixAMC 144 WR2.2, MixAMC 143 WR1.5, or SAX WR1.0 depending on the considered frequency band). For heterodyne measurements, the local oscillator is provided by a tunable frequency generator (Wiltron 68347B) while the intermediate frequency is measured on a spectrum analyzer (Agilent E4448A).

\section{Table of frequencies}

A table of all individual line frequencies measured with the hetereodyne mixers is given in Table~\ref{tab:tab1}.

\begingroup
\squeezetable
\begin{table*}[htp]
\caption{\label{tab:tab1} Lines measured with the heterodyne receivers.  The three lines noted with a $^\dagger $ sign where obtained by a Q-branch pump. All the other lines where obtained by an R-branch pump. (continued in Table III).}
\begin{tabular*}{\textwidth}{c|c|c|c|c|c|c|c|c|c|c|c}
QCL emission & $J_L$ & $J_U$ & $K$ & Lasing frequency  & Relative strength & QCL emission & $J_L$ & $J_U$ & $K$ & Lasing frequency  & Relative strength  \\
wavenumber (cm$^{-1}$) & & &  & (GHz) & (dB) & wavenumber (cm$^{-1}$) & & &  & (GHz) & (dB)  \\
\hline
1058.36 & 5 & 6 & 0 & 302.3208 &  -0.65 & 1069.9 & 13 & 14 & 7 & 704.1966 & -7.64 \\
1058.36 & 5 & 6 & 1 & 302.3148 &  -0.55  & 1071.3 & 14 & 15 & 0 & 755.1564 &  -6.32\\
1058.36 & 5 & 6 & 2 & 302.2962 &  -1.3 & 1071.3 & 14 & 15 & 1 & 755.1408 &  -7.68\\
1058.36 & 5 & 6 & 3 & 302.2651 &  0 & 1071.3 & 14 & 15 & 2 & 755.0949 &  -5.69\\
1058.36 & 5 & 6 & 4 & 302.2214 &  -10.62 & 1071.3 & 14 & 15 & 3 & 755.0174 &  0\\
1059.8 & 6 & 7 & 0 & 352.6872 &  -0.64 & 1071.3 & 14 & 15 & 4 & 754.9108 &  -3.2\\
1059.8 & 6 & 7 & 1 & 352.6798 &  -0.52 & 1071.3 & 14 & 15 & 5 & 754.7717 &  -3.96\\
1059.8 & 6 & 7 & 2 & 352.6581 &  -0.41 & 1071.3 & 14 & 15 & 6 & 754.6006 &  -2.12\\
1059.8 & 6 & 7 & 3 & 352.6222 &  0 & 1071.3 & 14 & 15 & 7 & 754.3977 &  -13.12\\
1059.8 & 6 & 7 & 4 & 352.5709 &  -3.31 & 1058.36 & 15 & 16 & 0 & 805.3878 &  -7.56\\
1059.8 & 6 & 7 & 5 & 352.5054 &  -7.91 & 1071.3 & 15 & 16 & 1 & 805.3698 &  -7.79\\
1059.8 & 6 & 7 & 6 & 352.4251 &  -9.2 & 1071.3 & 15 & 16 & 2 & 805.3212 &  -2.83\\
1061.3 & 7 & 8 & 0 & 403.0439 & -3.27  & 1071.3 & 15 & 16 & 3 & 805.2392 &  -11.90\\
1061.3 & 7 & 8 & 1 & 403.0351 & -4.25 & 1071.3 & 15 & 16 & 4 & 805.1262 &  0\\
1061.3 & 7 & 8 & 2 & 403.0103 &  -1.48 & 1071.3 & 15 & 16 & 5 & 804.9762 &  -0.6\\
1061.3 & 7 & 8 & 3 & 402.9692 &  0 & 1071.3 & 15 & 16 & 6 & 804.7952 &  -0.81\\
1061.3 & 7 & 8 & 4 & 402.9111 &  -1.4 & 1072.65 & 15 & 16 & 7 & 804.5807 &  -7\\
1061.3 & 7 & 8 & 5 & 402.8363 &  -8.6 & 1074 & 16 & 17 & 0 & 855.5951 &  -4.81\\
1061.3 & 7 & 8 & 6 & 402.7442 &  -1.67 & 1074 & 16 & 17 & 1 & 855.5774 &  -3.92\\
1062.85 & 8 & 9 & 0 & 453.389 &  -2.23 & 1074 & 16 & 17 & 2 & 855.5268 &  -2.7\\
1062.85 & 8 & 9 & 1 & 453.3803 &  -2.15 & 1074 & 16 & 17 & 3 & 855.4388 &  -0.32\\
1062.85 & 8 & 9 & 2 & 453.3514 &  -4.46 & 1074 & 16 & 17 & 4 & 855.317 &  -0.34\\
1062.85 & 8 & 9 & 3 & 453.3061 &  0 & 1074 & 16 & 17 & 5 & 855.1602 &  0\\
1062.85 & 8 & 9 & 4 & 453.2402 &  -1.51 & 1074 & 16 & 17 & 6 & 854.9668 &  -0.23\\
1062.85 & 8 & 9 & 5 & 453.1563 &  -6.12 & 1074 & 16 & 17 & 7 & 854.7402 &  -6.42\\
1062.85 & 8 & 9 & 6 & 453.0533 &  -0.73 & 1075.3 & 18 & 19 & 0 & 905.7797 &  -4.88\\
1064.3 & 9 & 9 & 0 & 503.7224 &  -6.19 & 1075.3 & 18 & 19 & 1 & 905.7626 &  -4.04\\
1064.3 & 9 & 9 & 1 & 503.7124 &  -8.67 & 1075.3 & 18 & 19 & 2 & 905.7085 &  -5.19\\
1064.3 & 9 & 9 & 2 & 503.6812 &  -5.83 & 1075.3 & 18 & 19 & 3 & 905.6162 &  0\\
1064.3 & 9 & 9 & 3 & 503.6299 &  0 & 1075.3 & 18 & 19 & 4 & 905.4883 &  -2.39\\
1064.3 & 9 & 9 & 4 & 503.5571 &  -3.75 & 1075.3 & 18 & 19 & 5 & 905.3233 &  -2.18\\
1064.3 & 9 & 9 & 5 & 503.4641 &  -4.67 & 1075.3 & 18 & 19 & 6 & 905.1175 &  -0.01\\
1064.3 & 9 & 9 & 6 & 503.3493 &  -7.69 & 1075.3 & 18 & 19 & 7 & 904.8773 &  -1.07\\
\end{tabular*}
\end{table*}
\endgroup

\begingroup
\squeezetable
\begin{table*}[h]
\caption{\label{tab:tab1} Lines measured with the heterodyne receivers.  The three lines noted with a $^\dagger $ sign where obtained by a Q-branch pump. All the other lines where obtained by an R-branch pump. (continued from Table II).}
\begin{tabular*}{\textwidth}{c|c|c|c|c|c|c|c|c|c|c|c}
QCL emission & $J_L$ & $J_U$ & $K$ & Lasing frequency  & Relative strength & QCL emission & $J_L$ & $J_U$ & $K$ & Lasing frequency  & Relative strength  \\
wavenumber (cm$^{-1}$) & & &  & (GHz) & (dB) & wavenumber (cm$^{-1}$) & & &  & (GHz) & (dB)  \\
\hline
1065.7 & 10 & 11 & 0 & 554.0426 &  -7.9 & 1076.66 & 19 & 20 & 0 & 955.9417 &  -7.58\\
1065.7 & 10 & 11 & 1 & 554.0307 &  -6.13 & 1076.66 & 19 & 20 & 1 & 955.9225 &  -5.59\\
1065.7 & 10 & 11 & 2 & 553.9969 &  -2.36 & 1076.66 & 19 & 20 & 2 & 955.8658 &  -9.81\\
1065.7 & 10 & 11 & 3 & 553.9400 &  0 & 1076.66 & 19 & 20 & 3 & 955.7675 &  -4.58\\
1065.7 & 10 & 11 & 4 & 553.8605 &  -2.96 & 1076.66 & 19 & 20 & 4 & 955.6328 &  -10.74\\
1065.7 & 10 & 11 & 5 & 553.7580 &  -13.28 & 1076.66 & 19 & 20 & 5 & 955.4578 &  -9.03\\
1065.7 & 10 & 11 & 6 & 553.6320 &  -2.06 & 1076.66 & 19 & 20 & 6 & 955.2444 &  0\\
1067.2 & 11 & 12 & 0 & 604.3463 &  -10.25 & 1077.75 & 20 & 21 & 0 & 1006.0747 &  0\\
1067.2 & 11 & 12 & 1 & 604.3346 &  -12.22 & 1077.75 & 20 & 21 & 1 & 1006.0574 &  -4.71\\
1067.2 & 11 & 12 & 2 & 604.2966 &  -6.29 & 1077.75 & 20 & 21 & 2 & 1005.9947 &  -4.56\\
1067.2 & 11 & 12 & 3 & 604.2352 &  0 & 1077.75 & 20 & 21 & 3 & 1005.8941 &  -2.18\\
1067.2 & 11 & 12 & 4 & 604.1482 &  -3.12 & 1077.75 & 20 & 21 & 4 & 1005.7516 &  -3.08\\
1067.2 & 11 & 12 & 5 & 604.0371 &  -11.25 & 1077.75 & 20 & 21 & 5 & 1005.5677 &  -0.65\\
1067.2 & 11 & 12 & 6 & 603.8997 &  -3.58 & 1077.75 & 20 & 21 & 6 & 1005.3424 &  -0.53\\
1047$^\dagger$ & 12 & 12 & 7 & 603.7364  &  -6.68 & 1079 & 21 & 22 & 0 & 1056.1816 &  -9.38\\
1047$^\dagger$ & 12 & 12 & 8 & 603.5475 &  -13.47 & 1079 & 21 & 22 & 1 & 1056.1617 &  -8.04\\
1047$^\dagger$ & 12 & 12 & 9 & 603.3327 &  -11.06 & 1079 & 21 & 22 & 2 & 1056.0986 &  -6.32\\
1068.69 & 12 & 13 & 0 & 654.6348 &  -2.17 & 1079 & 21 & 22 & 3 & 1055.9928 &  -1.93\\
1068.69 & 12 & 13 & 1 & 654.6214 &  -2.59 & 1079 & 21 & 22 & 4 & 1055.8418 &  -3.05\\
1068.69 & 12 & 13 & 2 & 654.5812 &  -3.12 & 1079 & 21 & 22 & 5 & 1055.6515 &  -4.58\\
1068.69 & 12 & 13 & 3 & 654.5147 &  -0.04 & 1079 & 21 & 22 & 6 & 1055.4141 &  0\\
1068.69 & 12 & 13 & 4 & 654.4207 &  -1.14 & 1079 & 21 & 22 & 7 & 1055.1358 &  -9.33\\
1068.69 & 12 & 13 & 5 & 654.3006 &  -7.01 & 1079 & 21 & 22 & 8 & 1054.811 &  -10.28\\
1068.69 & 12 & 13 & 6 & 654.1515 &  0 & 1080.3 & 22 & 23 & 0 & 1106.2619 &  -2.24\\
1069.9 & 13 & 14 & 0 & 704.9054 &  -5.98 & 1080.3 & 22 & 23 & 1 & 1106.2408 &  -1.57\\
1069.9 & 13 & 14 & 1 & 704.8910 &  -10.42 & 1080.3 & 22 & 23 & 2 & 1106.1735 &  -2.76\\
1069.9 & 13 & 14 & 2 & 704.8474 &  -2.11 & 1080.3 & 22 & 23 & 3 & 1106.0619 &  0\\
1069.9 & 13 & 14 & 3 & 704.7762 &  -3.12 & 1080.3 & 22 & 23 & 4 & 1105.9049 &  -7.71\\
1069.9 & 13 & 14 & 4 & 704.6753 &  -2.63 & 1080.3 & 22 & 23 & 5 & 1105.7057 &  -4.5\\
1069.9 & 13 & 14 & 5 & 704.5452 &  -1.42 & 1080.3 & 22 & 23 & 6 & 1105.4581 &  -2.25\\
1069.9 & 13 & 14 & 6 & 704.3856 &  0 & 1080.3 & 22 & 23 & 7 & 1105.1659 &  -6.11\\
\end{tabular*}
\end{table*}
\endgroup

\newpage

\section{Saturation of the infrared pump}

\begin{figure}[ht]
\centering
\includegraphics[width=0.85\linewidth]{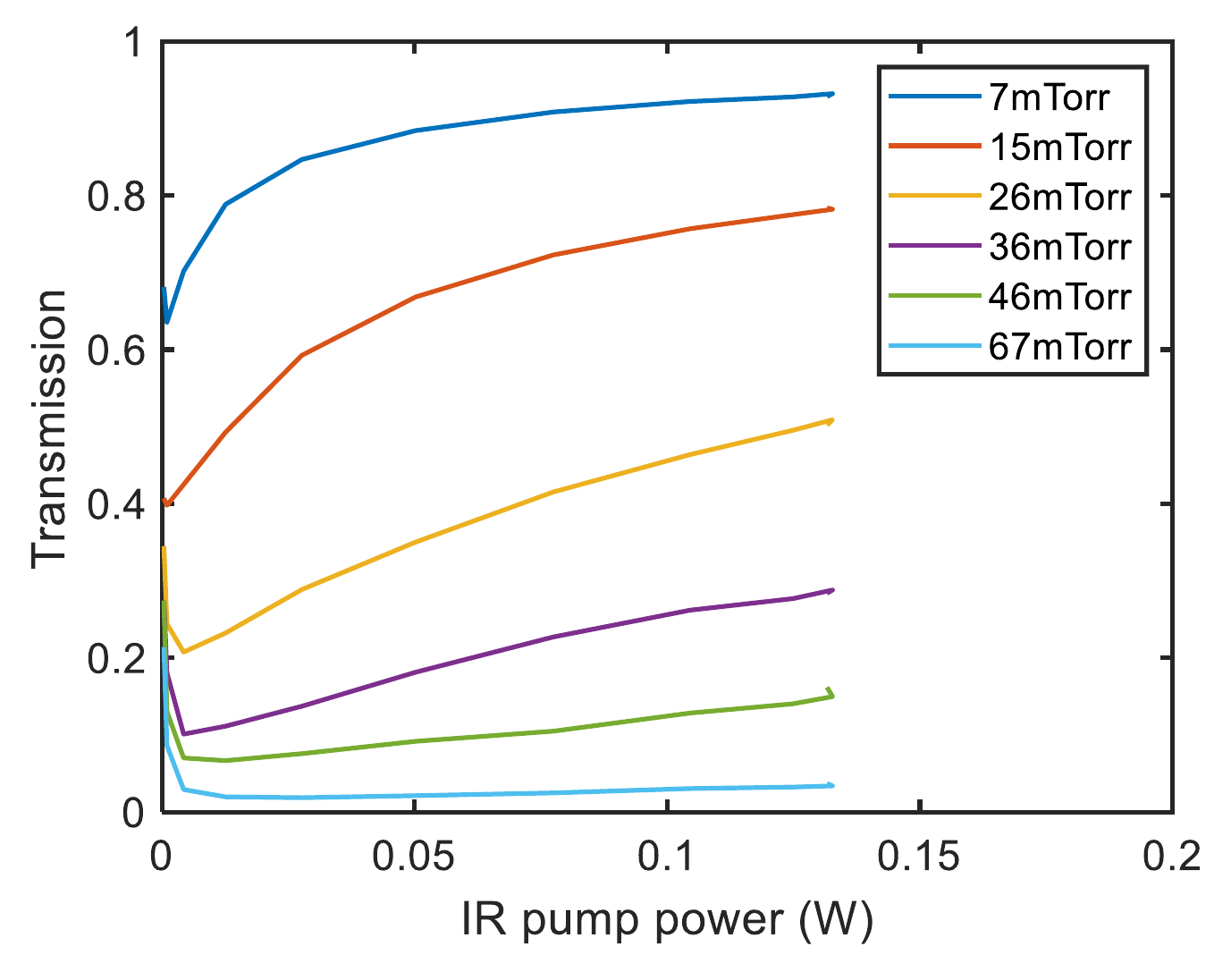}
\caption{Plot showing the transmission of the infrared pump through the laser cell (length 50 cm) as a function of the infrared power for different gas pressures. The infrared power is varied by means of a rotating polarizer. The QCL frequency was tuned to the $J=11 \rightarrow J=12, K=3$ transition of the vibrational mode described in the manuscript.}
\label{fig:saturation}
\end{figure}

When the back mirror of the cavity is removed and replaced by a vacuum-tight, anti-reflection coated Zinc Selenide window, one can study how much light is absorbed by the gas as a function of the infrared power during single pass through the cavity. The ratio of the transmitted power with gas in the cavity to the power transmitted without gas in the cavity is plotted as a function of the coupled infrared power in Fig~\ref{fig:saturation} for different pressures of the gas in a 50 cm long cell (same cell length as the one used in the main text). This plot shows the gas IR absorption strength strongly depends on the QCL input power, especially at low gas pressure.
This indicates a strong saturation of the infrared absorption that is due to the narrow linewidth of the QCL relative to the bandwidth of the IR transition (the Doppler broadened full width half maximum at 1050 cm$^{-1}$ is 67 MHz, the QCL linewidth is estimated around 1 MHz).

This strong pump saturation prevents our laser cavity for reaching higher output power at low pressures. This limitation can be overcome by using a better optimized cavity geometry and different pumping conditions, something that will be an objective of future work.

\section{Terahertz power vs pump detuning}

\begin{figure}[htbp]
\centering
\includegraphics[width=\linewidth]{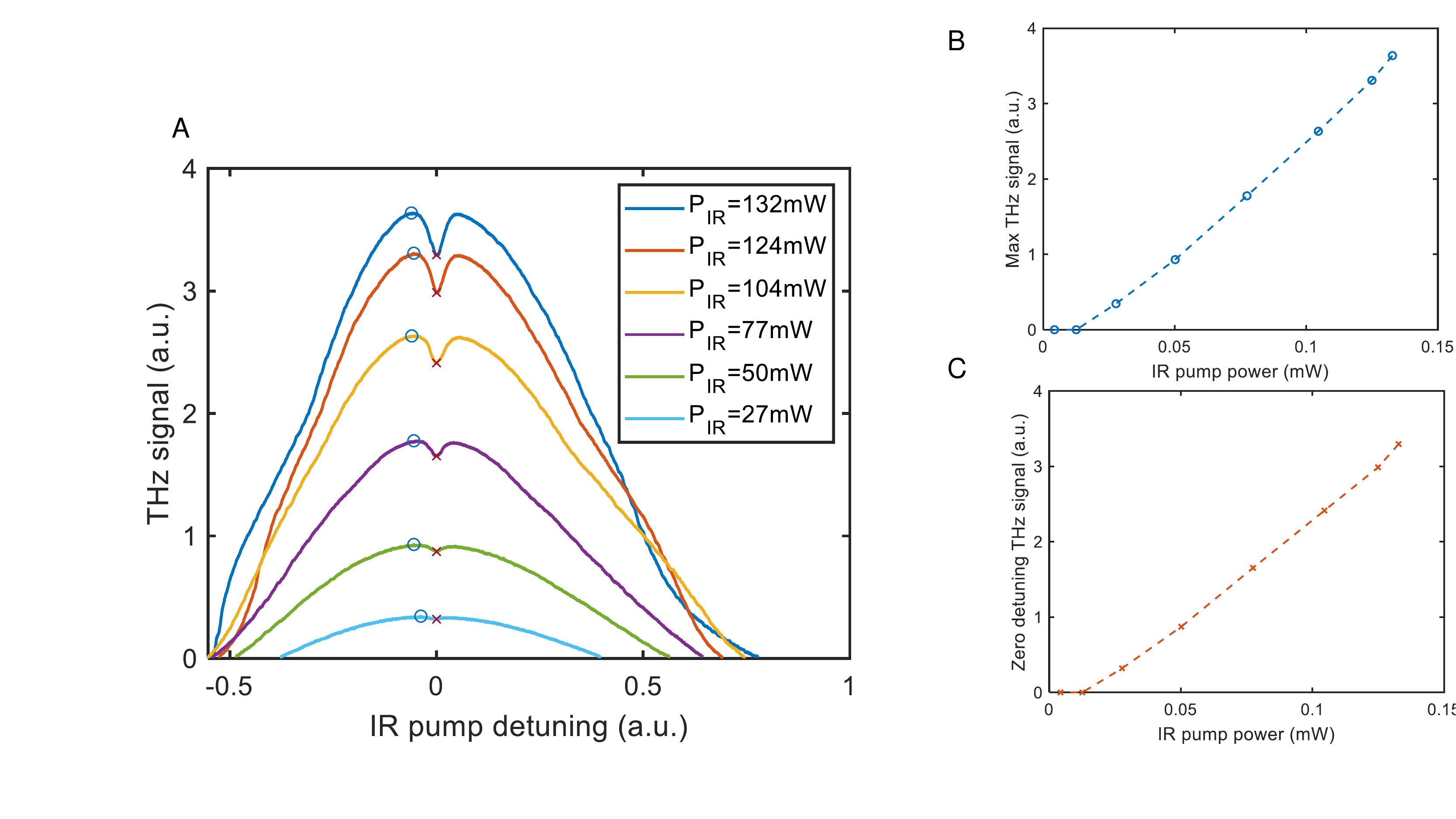}
\caption{(A) Plot showing the THz signal measured by the detector as a function of the pump detuning (obtained by a temperature scan of the pump QCL) for different pump powers. This corresponds to the lasing transition around 604 GHz between $J=12 \to 11$, K=3. Plot of the THz signal as a function of the pump power for the same transition (B) for the maximum signal and (C) for zero detuning.}
\label{fig:dip}
\end{figure}

The mid-infrared QCL can be continuously tuned around the peak absorption of the gas IR transition. Tuning of the QCL into the transition is typically achieved by tuning the laser temperature using small increments. Once laser emission is achieved, the QCL frequency can be continuously swept around the absorption line, and the THz signal can be monitored with the detector. Because the two counter propagating waves in the laser cavity encounter different gas velocity sub-classes, a Lamb dip-like feature is observed at the zero detuning point. 

The measured THz signal at 604 GHz ($J=12 \to 11, K=3$ transition) is plotted in Fig.~\ref{fig:dip}(A) as a function of the pump detuning. Figures~\ref{fig:dip}(B) and ~\ref{fig:dip}(C), respectively show the maximum signal and the signal at zero detuning 

\section{Laser linewidth and tuning range}

\begin{figure}[htbp]
\centering
\includegraphics[width=\linewidth]{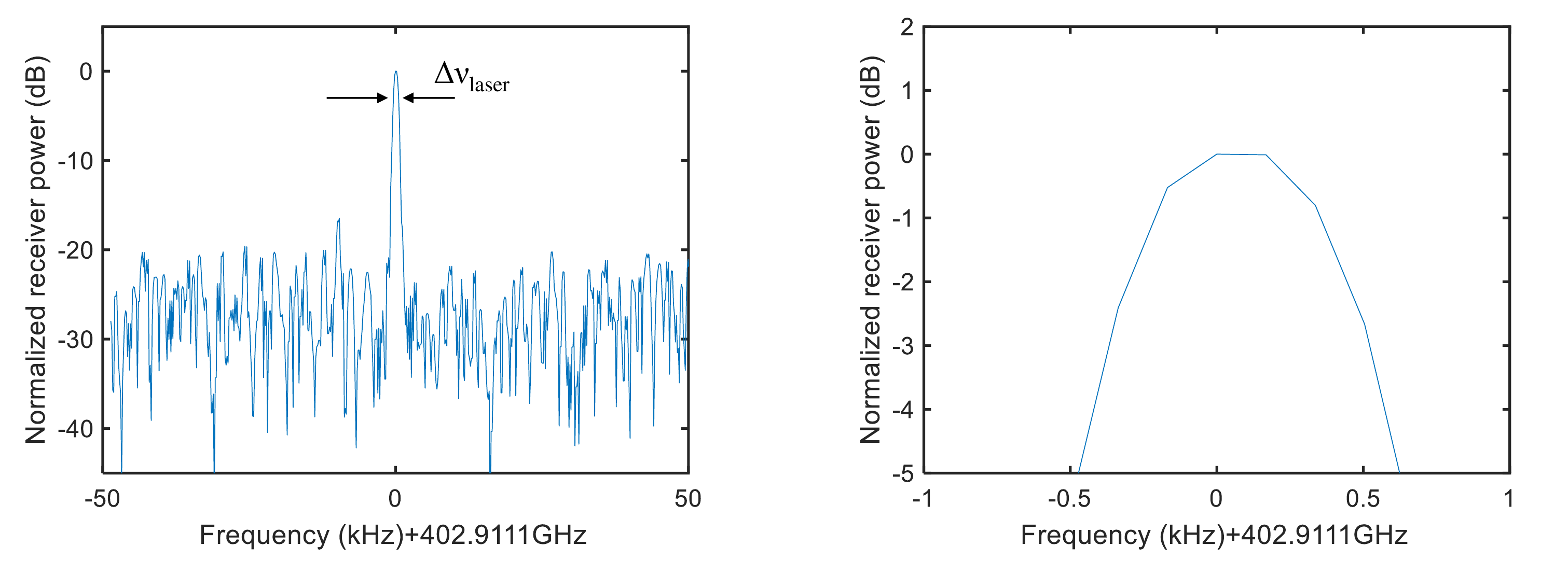}
\caption{Plot showing the measured line and a detail of the 3~dB linewidth for the lasing transition around 402.9 GHz, corresponding to the $J=8 \rightarrow 7, K=4$ molecular transition.}
\label{fig:linewidth}
\end{figure}

The laser linewidth can be measured experimentally, using the same heterodyne receiver that was used to recover each line frequency. In practice the linewidth stability seemed limited by the mechanical stability of the cavity. The smallest measured linewidth was recovered for one of the lowest frequency lasing lines measured (around 402.9 GHz, the $J=8 \rightarrow 7, K=4$ transition).
The recovered linewidth was less than 1 kHz, and a detailed view is shown in Fig.~\ref{fig:linewidth}.

A consequence of such narrow laser linewidth compared to other THz sources (such as frequency multiplier-based sources) is the very high brightness temperatures $T_b = I c^2/(2k_B \nu^2_\mathrm{THz} \Delta \nu) \approx 10^{15}$~K for laser radiance $I = 300$~W.m$^{-2}$.sr$^{-1}$), $k_B$ the Boltzmann constant, $c$ the speed of light, $\nu_{\mathrm{THz}}= 1$ THz the laser frequency and $\Delta \nu = 1$ kHz the laser linewidth.

\begin{figure}[htbp]
\centering
\includegraphics[width=\linewidth]{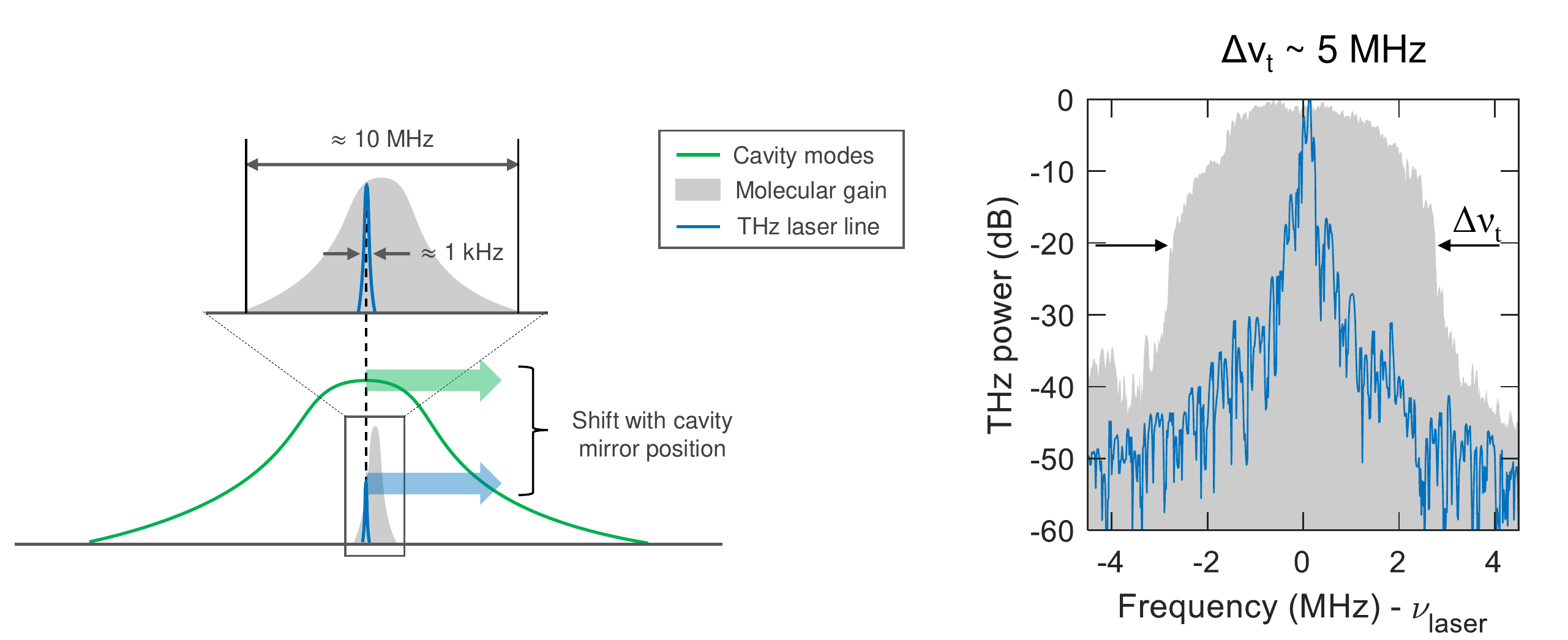}
\caption{(A) Schematic of the principle of cavity pulling: the cavity mode -- with a larger bandwidth than the molecular gain -- can be moved by tuning the cavity length. Lasing happens at the frequency with the lowest lasing threshold, that depends on the relative position of the gain bandwidth and of the cavity mode. (B) The blue curve shows a recovered frequency measured with the heterodyne mixer around 604 GHz ($J=12 \rightarrow 11, K=3$ transition) for a pump power of 150 mW and a gas pressure of 30 mTorr. The gray background corresponds to the envelope of all measured lasing frequencies obtained by tuning the cavity mirror by small steps.}
\label{fig:tuning}
\end{figure}

The laser line can be tuned within the molecular gain bandwidth by a cavity pulling effect, as shown in Fig.~\ref{fig:tuning}. 
Given the cavity length (50 cm), the free spectral range of the laser resonator is about 300 MHz.

At 604 GHz, the mirror travels by 250 \textmu m, corresponding to half the wavelength, as the cavity is scanned between two consecutive longitudinal Fabry-Perot modes. Therefore, at this frequency, a mirror positioning accuracy of 1 \textmu m results in a frequency accuracy of the cavity resonance of about 300/250=1.2 MHz.
Manual cavity length tuning was performed with a differential micrometer screw (Newport DM17-4) with a 100 nm sensitivity of the fine adjustment range, thus corresponding to 120 kHz frequency sensitivity for a 600 GHz laser emission. 
As the laser frequency increases, the frequency accuracy offered by the mirror tuning technique decreases due to a shorter laser wavelength.
A 1 THz, the sensitivity offered by the differential micrometer screw now corresponds to a 200 kHz frequency range.
The tuning range presented in Fig.~\ref{fig:tuning} was obtained by scanning the cavity length with a piezo-electric linear actuator (Newport picomotor) offering a 30 nm travel resolution.
Because of the larger pressure broadening coefficient of methyl fluoride, along with a lower lasing threshold in our current laser cell compared to the previous work with N$_2$O, the frequency-dependent tuning range achieved by cavity pulling (around 5 MHz near 604 GHz) was approximately 10 times larger than observed with N$_2$O. 

\section{Output terahertz power and lasing threshold}

\subsection{Simple model}

In the main text, the THz emission power has been estimated using a enhanced model~\cite{wang2021optimizing} that incorporates effects of pump saturation, dipole-dipole collisions and mutiple passes of the  infrared the pump in the laser cell. In comparison, the previously reported simple model 
expression~\cite{chevalier2019widely}  assumed a very low pressure scenario with the rate of unfavorable molecular dipole-dipole collisions much smaller than the rate of favorable hard collisions with cavity walls. This model also assumed a perfect match between the QCL emission linewidth with the gain medium absorption bandwidth and a single pass of the infrared pump. In such case, the THz emission power $P_{THz}$ at the emission frequency $\nu_{THz}$ can be estimated using the following equation~\cite{chevalier2019widely}:

\begin{equation}
P_{THz}=    \frac{T}{4} \frac{ \nu_{THz}}{\nu_{IR}} \frac{ \alpha_{IR}}{\alpha_{cell} } \left (P_{QCL} - P_{th} \right )
    \label{eq:pthz}
\end{equation}
where $\alpha_{IR}$ is the IR absorption of the gain medium at the pump frequency $\nu_{IR}$, $\alpha_{\mathrm{cell}}$ represents the total THz losses of the laser cavity, $T$ is the output coupling fraction, $P_{QCL}$ is the QCL pump power absorbed by the gain medium (which corresponds to the injected IR power minus the power absorbed by THz cavity walls), and $P_{th}$ is the threshold power as expressed below:
\begin{equation}
P_{th}=    \frac{h}{4 \pi} \frac{ \nu_{IR}}{\alpha_{IR}} (\alpha_{\mathrm{cell}} R_{\mathrm{cell}}) \frac{ u^2}{ | \langle J_L |  \mu |  J_U \rangle |^2} 
    \label{eq:pth}
\end{equation}
where $R_{\mathrm{cell}} $ is the radius of the THz cavity, $u$ is the average absolute molecular velocity, and $\langle J_L |  \mu |  J_U \rangle $ is the transition dipole matrix element of the lasing rotational transition.

The lasing transition dipole element can be expressed as a function of the molecular permanent dipole moment $\mu_0$ and the $J$ and $K$ quantum numbers~\cite{townes1975microwave}:

\begin{equation}
  | \langle J,K,V| \mu | J+1, K, V \rangle | ^ 2 = \mu_0^2 \frac{(J+1)^2 - K^2}{(J+1)(2J+3)}
    \label{eq:dipoleLasing}
\end{equation}

\subsection{Strength of the infrared transitions}

In the simple model given in equation~\ref{eq:pthz}, the THz power is expressed as a function of the absorption coefficient $\alpha_IR$, which can be calculated as follows:

\begin{equation}
\alpha_{IR} =    \frac{1-\exp(-\alpha_0 L) }{L}
    \label{eq:eqAlpha}
\end{equation}
where $L$ is the cavity length, and $\alpha_0$ is the IR absorption coefficient for a given roto-vibrational transition and is proportional to the square of the transition dipole matrix element~\cite{atkins2011molecular}:

\begin{equation}
\alpha_{0}((J,K,V) \rightarrow (J',K',V'))  \propto | \langle J,K,V| \mu | J', K', V'\rangle | ^ 2
    \label{eq:eq5}
\end{equation}
In the case where either the cell is short or the absorption strength is small, we have $\alpha_0 L\ll 1$, which leads to $\alpha_{IR} \approx \alpha_0 $.
The absorption coefficient $\alpha_0$ strongly depends on the population of energy levels, which is determined by the Boltzmann distribution and the molecular degeneracy.

The value of the dipole matrix element of an IR roto-vibrational transition depends on the type of the transition (P, Q, or R) and the initial and final quantum numbers~\cite{townes1975microwave}: 

\begin{equation}
  | \langle J-1,K,V+1| \mu | J, K, V \rangle | ^ 2_P \propto \mu_0^2 \frac{J^2-K^2}{J(2J+1)}
    \label{eq:dipoleP}
\end{equation}

\begin{equation}
  | \langle J,K,V+1| \mu | J, K, V \rangle | ^ 2_Q \propto \mu_0^2 \frac{K^2}{J(J+1)}
    \label{eq:dipoleQ}
\end{equation}

\begin{equation}
  | \langle J+1,K,V+1| \mu | J, K, V \rangle | ^ 2_R \propto \mu_0^2 \frac{(J+1)^2 - K^2}{(J+1)(2J+1)}
    \label{eq:dipoleR}
\end{equation}

From Eq.~\ref{eq:dipoleQ} and Eq.~\ref{eq:dipoleR} one can see that Q-branch transitions($\Delta J=0$) will be more efficient to pump large $K$ values (i.e. $K$ close to $J$), while P or R-branch transitions ($\Delta J=1$ or $-1$) will be more efficient to pump small $K$ values (i.e. $K$ close to zero), as demonstrated in Fig~.3(C) of the manuscript.

The molecular degeneracy increases proportionally to the value of $J$, but population decreases according to Boltzmann distribution. Degeneracy is highest at non-zero $K$ lines multiples of 3 ($K=3, 6$, etc.), leading to larger populations and, therefore, stronger IR absorption coefficients than neighboring lines, as can be seen in the Fig.~1(B) of the main text.

\subsection{Normalized threshold and power efficiency}

The relative IR absorption line strength used to estimate the relative QPML performance for a given $J,K$ is estimated by the following expression:

\begin{equation}
  L(J_{initial},J_{final},K)=  | \langle J_{final},K,V+1| \mu | J_{initial}, K, V \rangle | ^ 2 Pop(J_{initial},K),
    \label{eq:linestrength}
\end{equation}

where $Pop(J,K)$ is the the relative thermal population of the ground state energy level with quantum numbers J and K. This population factor is expressed in Eq.~\ref{eq:populationFactor}, as a function of the ground state energy level $F_{\mathrm{ground}}(J',K')$, calculated from Eq.~\ref{eq:eq1} using the rotational constants from the $\nu_3=0$ column of Table.~\ref{tab:tabConstants}, the energy level degeneracy $g(J,K)$, the gas temperature $T$ and the Boltzmann constant $k_B$.

\begin{equation}
  Pop(J,K)= \frac{\displaystyle g(J,K) \exp \left  (\frac{-F_{\mathrm{ground}}(J,K)}{k_BT} \right ) }{\displaystyle \sum_{J',K'} g(J',K') \exp \left (\frac{-F_{\mathrm{ground}}(J,K)}{k_BT} \right )}
    \label{eq:populationFactor}
\end{equation}

The energy level degeneracy factor $g(J,K)$ can be expressed as follows:

\begin{equation}
  g(J,K)=
    \begin{cases}
      2J+1, & \text{if}\ K=0 \\
      4 \times (2J+1), & \text{if}\ K>0 \  \text{and} \ K \  \text{is a multiple of 3}\\
      2 \times (2J+1), & \text{otherwise}
    \end{cases}
    \label{eq:degeneracyFactor}
\end{equation}

The normalized power efficiency is then calculated as follows for Q and R branch, respectively: the relative IR absorption line strength is multiplied by the relative lasing frequency as a function of the pumped quantum number $J_{L}$ and $K$

\begin{equation}
\begin{split}
  \eta_Q(J_{L},K) &= ( J_{L}) L(J_{L},J_{L},K) \\
  &= ( J_{L}) | \langle J_{L},K,V+1| \mu | J_{L}, K, V \rangle | ^ 2 Pop(J_{L},K)
    \label{eq:QbranchPE}
    \end{split}
\end{equation}

\begin{equation}
\begin{split}
  \eta_R(J_{L},K) &= ( J_{L}+1) L(J_{L},J_{L}+1,K) \\ 
  &= ( J_{L}+1) | \langle J_{L}+1,K,V+1| \mu | J_{L}, K, V \rangle | ^ 2 Pop(J_{L},K) 
    \label{eq:RbranchPE}
    \end{split}
\end{equation}

The normalized lasing threshold is calculated for Q and R branch respectively, by taking the inverse of the line strength and the lasing matrix element as a function of the pumped quantum number $J_{L}$ and $K$

\begin{equation}
\begin{split}
  Th_Q(J_{L},K) &= \frac{1} { | \langle J,K,V| \mu | J+1, K, V \rangle | ^ 2 L(J_{L},J_{L},K)} \\
  &=  \frac{1} { | \langle J,K,V| \mu | J+1, K, V \rangle | ^ 2 | \langle J_{L},K,V+1| \mu | J_{L}, K, V \rangle | ^ 2 Pop(J_{L},K)  }
    \label{eq:Qbranchth}
    \end{split}
\end{equation}

\begin{equation}
\begin{split}
  Th_R(J_{L},K) &= \frac{1} { | \langle J_L,K,V| \mu | J_L+1, K, V \rangle | ^ 2 L(J_{L},J_{L}+1,K) } \\ 
  &=  \frac{1} { | \langle J_L,K,V| \mu | J_L+1, K, V \rangle | ^ 2  | \langle J_{L}+1,K,V+1| \mu | J_{L}, K, V \rangle | ^ 2 Pop(J_{L},K)}
    \label{eq:Rbranchth}
    \end{split}
\end{equation}

\bibliography{references}